# HOW EFFECTUAL WILL YOU BE? DEVELOPMENT AND VALIDATION OF A SCALE IN HIGHER EDUCATION

Alicia Martín-Navarro [a], José Aurelio Medina-Garrido [b,*], Félix Velicia-Martín [c]

[a] *Universidad de Sevilla. INDESS, Universidad de Cadiz, Jerez de La Frontera, Cadiz, Spain*
[b] *INDESS, Universidad de Cadiz, Jerez de La Frontera, Cadiz, Spain*
[c] *Universidad de Sevilla, Sevilla, Spain*
[*] *Corresponding author*

This is the preprint version accepted for publication in the "International Journal of Management Education". The final published version can be found at: https://doi.org/10.1016/j.ijme.2021.100547 We acknowledge that Elsevier Sci LTD holds the copyright of the final version of this work. Please, cite this paper in this way:

Martín-Navarro, A., Medina-Garrido, J. A., & Velicia-Martín, F. (2021). How effectual will you be? Development and validation of a scale in higher education. International Journal of Management Education, 19(3), 100547.

**ABSTRACT**

The literature on effectual theory offers validated scales to measure effectual or causal logic in entrepreneurs' decision-making. However, there are no adequate scales to assess in advance the effectual or causal propensity of people with an entrepreneurial intention before the creation of their companies. We aim to determine the validity and reliability of an instrument to measure that propensity by first analysing those works that provide recognised validated scales with which to measure the effectual or causal logic in people who have already started up companies. Then, considering these scales, we designed a scale to evaluate the effectual or causal propensity in people who had not yet started up companies using a sample of 230 final-year business administration students to verify its reliability and validity. The validated scale has theoretical implications for the literature on potential entrepreneurship and entrepreneurial intention and practical implications for promoters of entrepreneurship who need to orient the behaviour of entrepreneurs, entrepreneurs of established businesses who want to implement a specific strategic orientation, entrepreneurs who want to evaluate the effectual propensity of their potential partners and workers, and academic institutions interested in orienting the entrepreneurial potential of their students.

**INTRODUCTION**

Economic activity is driven by entrepreneurship, making it an essential tool for the creation of jobs and the generation of wealth (Oosterbeek, van Praag, & Ijsselstein, 2010). The need to overcome the economic problems of the last decade has stimulated the idea that the self-employed should take a more active role, not only out of necessity, but because self employment can lead to innovation, employment, and economic and social development (Montañés-Del-Río & Medina-Garrido, 2020; Sánchez García, Ward, Hernández, & Florez, 2017).

Studying the behaviour of the self-employed can explain how they further develop the entrepreneurial process, making it possible to understand how they create and make new companies grow and how they generate economic growth (Fisher, 2012). In studies of the entrepreneurial phenomenon, the explanations for the behaviour of entrepreneurs that effectuation theory offers stand out (Sarasvathy, 2001), and research in the field of entrepreneurship increasingly uses this theory as a conceptual basis. Despite effectuation is still a new line in entrepreneurship research, this theory has developed rapidly in the last years (Matalamäki, 2017).

In Sarasvathy's (2001) effectuation theory, the concepts of effectuation and causation provide a framework for studying entrepreneurial processes in environments with different levels of uncertainty. Effectuation refers to processes that are carried out thanks to the possible results that entrepreneurs can obtain from the available resources. However, the concept of causation has a predictive basis. In this case, starting a business involves selecting the means necessary to create the desired result. In short, the difference concerning the entrepreneurial process is that for causal logic, opportunities are created, whereas for effectual logic, opportunities are discovered (Vaghely & Julien, 2010).

The effectual versus causal logic of entrepreneurs has been studied in the academic literature primarily by examining the behaviour of those people who have started businesses (Brettel, Mauer, Engelen, & Küpper, 2012; Guo, Cai, & Zhang, 2016; Schmidt & Heidenreich, 2014); that is, from the evidence of the entrepreneurs' actions, which are often guided by the entrepreneur's prior social identity (Alsos et al., 2016). However, few studies have delved into the individual's propensity towards these logics when the potential entrepreneur has not yet acted. Given this gap, the study of the propensity towards an effectual or causal behaviour gives added value to the literature on entrepreneurship. The only partially similar framework found in the literature is the work of Werhahn et al. (2015). These authors analyse the effectual orientation of managers and workers in the corporate context finding that the strategic orientation of the company towards an effectual or causal logic favours an organisational culture that affects workers' behaviour.

The detected gap presents a dilemma in terms of whether it is possible to measure the causal or effectual propensity of a person before they start a business. This research aims to elaborate and validate a measurement tool that estimates an individual's propensity towards an effectual or causal behaviour when they have not yet started a business.

This scale would have important implications for the literature on potential entrepreneurship and entrepreneurial intention. It would also be useful for promotors of entrepreneurship, entrepreneurs who wish to implement a specific strategic orientation, entrepreneurs who wish to measure the effectual propensity of their potential partners and workers, and academic institutions interested in developing and orienting the entrepreneurial potential of their students.

To achieve this objective, we first analysed the scales accepted in the literature to measure the effectual or causal logic of entrepreneurs. Then we designed a questionnaire to measure the propensity towards these two logics of people who had not yet started a business, using a sample of university students in their final year of studying for a business administration degree at the University of Cadiz and the University of Seville (Spain). Subsequently, we verified the reliability and validity of the scale. The discussion section analyses the results and presents the theoretical and practical implications of the validated scale for measuring the effectual or causal propensity of potential entrepreneurs.

**BACKGROUND**

Entrepreneurs continually make decisions regarding business idea improvements, the creation or identification of new market niches, the resolution of technical problems, and the recruitment and selection of key personnel in the company, among others (Davidsson & Klofsten, 2003). Effectuation theory (Sarasvathy, 2001) differentiates between effectual logic and causal logic in the decisions that entrepreneurs make under situations of uncertainty and risk. Under a logic oriented to causality, entrepreneurs recognise, develop, and evaluate opportunities. Then they identify the resources needed to exploit them and evaluate their feasibility. By contrast, the effectual vision of the entrepreneurial process begins with the resources available to entrepreneurs and the opportunities they create with them (Perry, Chandler, & Markova, 2012). The causal logic assumes that markets and opportunities are objective and pre-existing, the business process is linear and unidirectional, and the known outcome is the desired goal. On the other hand, the effectual logic presumes that markets and opportunities are subjective and limited only by the imagination of the entrepreneur, that the entrepreneurial process is dynamic, interactive, and non-linear, and it considers a set of resources that is already available (Dutta & Thornhill, 2014). For this reason, expert entrepreneurs use effectual logic to create new markets (Dew, Read, Sarasvathy, & Wiltbank, 2011)

Either approach, effectuation or causation, may be required at different times in the evolution of the company. An entrepreneur can use both logics indistinctly, depending on the uncertainty of the circumstances (Gabrielsson & Politis, 2011). In other words, the processes of effectuation and causation can coexist and are shaped in a specific way in the different phases of the life cycle of the company (Matalamäki, Vuorinen, Varamäki, & Sorama, 2017). Therefore, it is common for experienced decision-makers to possess an accumulated knowledge of trial-and-error decision-making, while new entrepreneurs prefer to take the approach of developing a business plan for the business they wish to create (Dew, Read, Sarasvathy, & Wiltbank, 2009), which they learned from the entrepreneurship programmes they attended in educational settings.

**Causation**

The specific environment in which entrepreneurs operate is critical, as it will influence their perception of what is preferable and most effective. Causal logic is valuable in situations where the existing market is definable and measurable (Gabrielsson & Politis, 2011). Causation is a model for decision-making and problem-solving based on the logic of prediction. The entrepreneurs' knowledge of the available means and the output they wish to obtain drive the selection of resources. It is assumed that the market exists independently of entrepreneurs and that their main task is to capture the largest possible share of that market. Entrepreneurs try to achieve this objective by planning for and gathering the necessary information to see how the strategies materialise according to the plan and also by identifying the possible causes for the result differing from the plan (Sarasvathy, 2001; Sarasvathy & Dew, 2005).

**Effectuation**

In the creation of new companies, entrepreneurs who follow an effectual approach often start the process with the sole aspiration of creating a company. The experience of starting and running new businesses makes successful entrepreneurs more inclined towards effectual logic in their decision-making (Baron, 2009). Since they make decisions and observe the results of those decisions, they use the new information to change course. As the future is unpredictable for them, their effectual logic makes them address the market from different perspectives before deciding on a business model (Sarasvathy, 2001). According to Sarasvathy (2001, 2008), the effectual logic considers five dimensions, namely means, partnership, affordable loss, contingency, and control. Although other authors have considered only four of the five dimensions, which have not always been the same four dimensions (Brettel et al., 2012; Chandler, DeTienne, McKelvie, & Mumford, 2011; Perry et al., 2012), this work will consider the original five dimensions proposed by Sarasvathy.

**Means orientation**

Sarasvathy (2001) argues that in the effectual process, entrepreneurs use the resources (means) they already have and select a possible output that they can create with those resources. The resources fall into three different categories: (1) who the person is — that is, the entrepreneur's traits, preferences, and skills; (2) what the person knows — that is, the entrepreneur's knowledge; and (3) who the person knows, which results from the entrepreneur's social relationships. Means orientation implies that the entrepreneur uses all available means to create a business. Thus, effectual entrepreneurs bring their knowledge, social networks, experience, and skills to their project (Werhahn et al., 2015).

**Partnership orientation**

Social networks contribute significantly to innovation (Granovetter, 1973) and uncertainty management (Krackhardt, 1992). Building alliances can help to control certain situations (Wiltbank, Dew, Read, & Sarasvathy, 2006). Social interaction plays an important role in the effectuation process (Fischer & Reuber, 2011). The effectual entrepreneur seeks out and identifies partners with whom to start a business and commit in order to achieve a mutual benefit (Read, Song, & Smit, 2009; Rese, 2006). This type of entrepreneur seeks the joint creation of new products or services or any other type of cooperation to take advantage of new business opportunities. Effectual entrepreneurs looking for these opportunities are aware that they depend mainly on engagements with others to share knowledge, ideas, networks, money, or time. (Werhahn et al., 2015). In this way, when partners commit and work together, they better control the future of their business and reduce their uncertainty (Sarasvathy, 2001).

**Affordable loss orientation**

According to Sarasvathy (2001), entrepreneurs who have an effectual orientation consider affordable losses to be more important than expected returns. This type of entrepreneur determines how much loss can be assumed and focuses on experimenting with as many strategies as possible to achieve success. Affordable loss becomes an important criterion on which entrepreneurs base their decisions. This way, entrepreneurs reject projects that cost more than they can afford to lose in favour of more affordable projects. Therefore, following this approach, each new venture would be seen as a project where the losses are under control. Additional resources would be incorporated only if the results justified it.

**Contingency orientation**

Companies have to deal every day with unexpected situations that are challenging to predict, particularly when they operate in uncertain environments (Hitt, Keats, & DeMarie, 1998). These surprises, whether positive or negative, give rise to setbacks. Effectual logic embraces these setbacks to pursue new business or market opportunities (Sarasvathy, 2008). The effectual entrepreneur does not see setbacks as obstacles to overcome, but as new resources and opportunities to exploit (Harmeling, 2011). This type of entrepreneur excels at pivoting to take advantage of unanticipated events (Sarasvathy, 2001). The individual must be willing to adapt in order to capitalise on these contingencies when faced with new information, means, or surprises (Read et al., 2009). Thus, a contingency orientation relates to the ability to make fast, creative, proactive, and effective changes. In other words, the effectual entrepreneur tries to take advantage of unforeseen events continuously and as advantageously as possible. The contingency orientation addresses the question of what the entrepreneur will decide to do about the environment and not how the entrepreneur will adapt to that environment (Werhahn et al., 2015).

**Control orientation**

Like organisations, individuals want to ensure favourable results (Wiltbank, Read, Dew, & Sarasvathy, 2009). The concept of control has particular relevance when the future is uncertain, as it seeks to exert some kind of influence on that future. This control implies believing that actors can proactively create or co-create their environment (Sarasvathy, 2001). From this perspective, the behaviour of companies and markets depends on individuals who do not conceive the future as an inevitable result of economic or technological forces. Thus, effectual entrepreneurs rarely see an opportunity as being out of their control, and what they do is work to discover and create opportunities (Dew, Read, Sarasvathy, & Wiltbank, 2008). Control orientation implies that individuals perceive the environment as endogenous and therefore controllable. In this way, the individual is motivated to influence socio-economic trends and to create new markets. The control-oriented individual can face new challenges successfully, especially in uncertain situations (Werhahn et al., 2015).

**EFFECTUAL PROPENSITY**

The existing literature has widely validated the measurement of effectual versus causal behaviour for the case of the entrepreneur who has already demonstrated such a behaviour (Brettel et al., 2012; Chandler et al., 2011). However, there is a gap in the literature if the aim is to measure the propensity of individuals to develop effectual or causal behaviours before these behaviours effectively take place. Being able to measure this propensity could have important practical implications. It would allow public, economic, and educational decision-makers to guide and refine these behaviours before they manifest. In this way, the potential entrepreneur will have the capacity to choose to develop an effectual or causal behaviour (or take a mixed position between the two behaviours) depending on the environmental circumstances (Futterer, Schmidt, & Heidenreich, 2018). Adequately training potential entrepreneurs will allow them to consciously consider each of the attributes of both orientations as they develop their entrepreneurial behaviour.

As noted above, the literature does not provide a validated scale to measure this effectual or causal propensity. In the methodology section, we adapted Werhahn et al.'s (2015) effectual orientation scale, which they validated in a corporate context. It was useful for this paper because it captures how an effectual strategic orientation at the corporate level can influence the effectual propensity of employees. Moreover, we validated a scale to measure the "causal propensity" of individuals that was adapted from the work of Chandler et al. (2011) and Gabrielsson and Politis (2011).

## METHODOLOGY

### Questionnaire

The questionnaire is one of the methods most widely used by researchers in the field of entrepreneurship (Brettel et al., 2012; Dutta, Gwebu, & Wang, 2015; Estrada, Cruz, Jover, & Gras, 2018; Guo et al., 2016). We adapted a questionnaire from the work of Werhahn et al. (2015), Chandler et al. (2011), and Gabrielsson and Politis (2011) to measure effectual and causal propensity in individuals who had not yet effectively developed such types of behaviour. On the one hand, the Werhahn et al. (2015) scale provided a measure of potential effectual behaviour adjusted to the original principles conceptualised by Sarasvathy and her colleagues (Dew et al., 2009; Sarasvathy, 2001) On the other hand, the scales of Chandler et al. (2011), and Gabrielsson and Politis (2011) completed the measurement of individuals' effectual orientation with the measurement of their causal orientation. When the questionnaire used is an adaptation of one developed by other authors, given that the proposed new instrument will not fully reflect the consistency of the original works, it is essential to re-establish its validity (Mendoza & Garza, 2009).

After finalising the initial design, a panel of entrepreneurship experts reviewed the questionnaire. Six researchers, six entrepreneurs and six entrepreneurship promoters participated in this panel. We contacted these experts by telephone or videoconference. Subsequently, a group of 34 students who met the expected profile reviewed the questionnaire. As a result, we produced a well-designed questionnaire that ensured that respondents fully understood the questions.

Once the refinement process was complete, we structured the definitive questionnaire into two differentiated sections (Lee, Tsao, & Chang, 2015). The first section included five demographic items: gender, age, nationality, employment status, and whether the student came from an entrepreneurial family. The second section included six categories that added up to a total of 25 questions to determine the effectual or causal propensity of the respondent.

**Sample and data collection**

To validate the new scale of measurement, we selected a sample of university students in their final year of studying for a business administration degree at the University of Cadiz and the University of Seville (Spain). Although they had not yet demonstrated effectual or causal behaviour, they could show a propensity towards the type of behaviour they would embody if they created a business. It was essential to study this propensity before they received specific training in entrepreneurship since this training would condition them. Therefore, the responses to the questionnaire were gathered in the first semester of the academic year before the students took a course on "business creation" in which they would acquire the knowledge and necessary skills to carry out a business plan. Data collection took place in October and November 2018. We created an online questionnaire using Google's Forms application (Hariguna, Lai, & Chen, 2016; Jiang & Wu, 2016; Lian, 2017), which allows the user to access the survey through a web link. This tool funnels the data directly into a spreadsheet, which facilitates subsequent statistical processing.

We obtained a sample of 230 completed questionnaires from a total of 463 students enrolled, obtaining a response rate of 49.67%. Of those surveyed, 58.3% were women. Furthermore, 86.1% of those surveyed were between 18 and 24 years old, 11.7% were between 25 and 30 years old, and the rest were more than 30 years old. Students from the University of Seville accounted for 58.3% of the respondents, while the remaining 41.7% were from the University of Cadiz. Moreover, 67.4% of those surveyed were pursuing a business administration degree, while 32.6% were pursuing a double degree in business administration and law. Of those surveyed, 36.1% had parents who had owned or currently owned a business, 43.5% had worked or were working as employees (as compared with 48.3% who had never worked), and the rest either had been or were self-employed or were both self-employed and employed. Table 1 shows the descriptive information on the sample.

Table 1. Descriptive information on the sample

| | |
|---|---|
| Sample | 230 (Response rate: 49.67%) |
| Gender | Female: 58.3%; Male 41.7% |
| Age | 18-24: 86.1% |
| | 25-30: 11.7% |
| | 30-: 2.2% |
| Parents who own/had owned a business | 36.1% |
| Working experience | As employees: 43.5% |
| | As self-employed: 8.2% |
| | Never worked: 48.3% |

**Measures**

Following Sarasvathy (2001) and Gabrielsson and Politis (2011), we did not consider effectual propensity and causal propensity to be two extremes on the same scale. Instead, we divided the logic of the respondents' decision-making into two different variables. Thus, the construct for effectual propensity was different and separate from the construct for causal propensity, making it possible to detect any combination of the two approaches.

On the one hand, we adapted the items for the construct "effectual propensity" from the work of Werhahn et al. (2015), which contemplates five dimensions: (1) Means Orientation; (2) Partnership Orientation; (3) Affordable Loss Orientation; (4) Contingency Orientation; and (5) Control Orientation. Means Orientation is the tendency of individuals to use the resources they already have and select a possible outcome that can be achieved with those resources. We measured this variable using three items. Partnership Orientation is the propensity of individuals to seek out partners to develop their projects, committing themselves to achieve mutual benefit. Four items measured this variable. Affordable Loss Orientation implies the tendency of individuals to make decisions by limiting their losses rather than focusing on expected returns. This variable was measured using three items. Contingency Orientation is the tendency of individuals to accept setbacks generated in uncertain environments as opportunities to be exploited. Four items measured this variable. Control Orientation is the tendency of individuals to proactively create their environment, especially in situations of uncertainty, by working to discover and create new opportunities. We measured this variable using four items.

On the other hand, the Causal Propensity of individuals is the tendency to make decisions and solve problems based on the selection of the necessary resources according to the available means and the desired outcome. Seven items measured Causal Propensity. Items one, three, and six were adapted from the work of Chandler et al. (2011). Items two, four, five, and seven were adapted from the work of Gabrielsson and Politis (2011). Following Dittrich et al. (2005), the above items were measured on a seven-point Likert scale, with one corresponding to "strongly disagree" and seven to "strongly agree". Table 2 shows all the items.

**RESULTS**

We used SmartPLS software to carry out the analysis of the measurement model. It allows for analysing the relationships between latent variables and their indicators. This analysis shows whether all the indicators represent their corresponding construct or whether some of them need to be removed. SmartPLS evaluates the measurement of variables based on individual reliability, construct reliability, discriminant validity, and convergent validity. This analysis ensures that the indicators are good at representing their corresponding variable. Reliability ensures that the measurement produces consistent results, and validity ensures that the indicators of a construct measure only their construct and not another.

First, the individual reliability of each item was analysed by simple correlations of the indicators with their respective construct. According to Carmines and Zeller (1979), loads must be $\lambda \geq 0.707$ to accept an indicator. However, some researchers consider that this rule of thumb ($\lambda \geq 0.707$) should not be so rigid in the initial stages of scale development (Hair Jr, Black, Babin, & Anderson, 2014). Hair et al. (2014) establish that indicators with loads between 0.4 and 0.7 could be removed from a scale if their suppression leads to an increase in the mean extracted variance (AVE) or composite reliability (CR) above the suggested threshold value (AVE = 0.5; CR = 0.7). In any case, researchers should eliminate indicators with very low loads (i.e., $\leq 0.4$) (Hair Jr, Ringle, & Sarstedt, 2011). In this study, there were no indicators with loads below 0.4, although there were indicators with loads between 0.4 and 0.7 (see Table 2) that could be eliminated at a later stage.

Second we analysed the reliability of the construct to see how rigorously the indicators measured the same latent variable. For this purpose, the measures corresponding to the composite reliability (Werts, Linn, & Jöreskog, 1974) should be greater than 0.8 (Nunnally & Bernstein, 1995). Dijkstra–Henseler's rho ($\rho A$) is another indicator to determine the reliability of the construct. Its value must be above 0.7 (Dijkstra & Henseler, 2015). In this study, all indicators met this requirement except for the Dijkstra–Henseler indicator in the construct "affordable losses", since its result was 0.691, which did not exceed the established value of 0.7. However, we did not remove the indicator since the value was very close to 0.7, it met the composite reliability threshold (which is a measure of greater acceptance than the indicator $\rho A$), and the loads of the indicators that make up this variable were quite high (see Table 2).

**Table 2**. Convergent reliability and validity before eliminating indicators

| Constructs/Indicators | Loads | Composite Reliability | ρA | AVE |
|---|---|---|---|---|
| **Means** | | 0.868 | 0.773 | 0.686 |
| I use my personal knowledge and experience in the best possible way. | 0.825 | | | |
| I pursue those initiatives for which I have great motivation and interest. | 0.814 | | | |
| I pursue those initiatives for which I personally have the relevant competencies. | 0.845 | | | |
| **Partnership** | | 0.812 | 0.708 | 0.524 |
| When I work with others, I aim to ensure that gains and risks are shared fairly. | 0.608 | | | |
| I approach potential partners very early on in order to jointly co-create new things. | 0.811 | | | |
| I enter into relationships with partners who are willing to commit (e.g. invest time) from the onset. | 0.816 | | | |
| When new actors appear in my environment, I perceive them as potential partners. | 0.634 | | | |
| **Affordable Loss** | | 0.819 | 0.691 | 0.601 |
| I would only invest in a business what I can afford to lose. | 0.823 | | | |
| In a business, I would try to limit the potential loss of initiatives to an acceptable degree, although it could be that by investing more, I would finally obtain benefits. | 0.726 | | | |
| I would only invest in my business if the loss of the investment would not ruin the company, although it could be that by investing more, I would finally obtain benefits. | 0.775 | | | |
| **Contingency** | | 0.880 | 0.826 | 0.649 |
| I regard surprises to be new opportunities that I could take advantage of. | 0.784 | | | |
| I exploit contingencies as effectively as possible. | 0.888 | | | |
| When I have new information, I try to take advantage of it. | 0.737 | | | |
| I use setbacks as new opportunities to take advantage of. | 0.805 | | | |
| **Control** | | 0.835 | 0.744 | 0.561 |
| I attempt to shape the environment I operate in. | 0.810 | | | |
| I attempt to proactively design my environment with others. | 0.816 | | | |
| In a business, we must attempt to create with others new needs for the market. | 0.603 | | | |
| I attempt to influence trends. | 0.748 | | | |
| **Causation** | | 0.835 | 0.786 | 0.424 |
| I usually design a long-term plan to organise myself in my tasks. | 0.689 | | | |
| I prefer to have predetermined goals and to strive to achieve the results of these goals. | 0.748 | | | |
| I analyse long-run opportunities and select what I think would provide the best returns. | 0.732 | | | |
| I try to avoid uncertain situations to the greatest possible extent. | 0.488 | | | |
| When I set goals to achieve, I analyse my competitors in depth. | 0.624 | | | |
| I usually implement control processes to make sure I meet the objectives. | 0.640 | | | |
| I think my relationships with those who can influence my future should be long term and goal oriented. | 0.600 | | | |

Note: ρA: Dijkstra–Henseler; AVE: Average extracted variance.

Subsequently, convergent validity was studied to verify that the indicators represent a single underlying construct. The average extracted variance (AVE) was used as a measure of this validity, requiring values greater than 0.5 (Fornell & Larcker, 1981). The latent variable "Causation" explained less than 50% of the variance of the indicators that comprised it (Table 2). We removed the following two items to solve this problem: "I try to avoid uncertain situations to the greatest possible extent" and "When I set goals to achieve, I analyse my competitors in depth". As a result, the AVE reached an acceptable score above the recommended 0.5.

Finally, we analysed the discriminatory validity before removing the previous items. The discriminatory validity is the extent to which a given construct is different from other constructs. Following Henseler et al. (2016), there is discriminant validity when the heterotrait-monotrait ratio of correlations (HTMT) has values below 0.85 (Gold, Malhotra, & Segars, 2001). Table 3 shows that all values were below this threshold, so we did not need to remove any other indicators.

**Table 3**. Discriminant validity before eliminating indicators

|  | Means | Partnership | Affordable Loss | Contingency | Control | Causation |
|---|---|---|---|---|---|---|
| **Means** |  |  |  |  |  |  |
| **Partnership** | 0.713 |  |  |  |  |  |
| **Affordable Loss** | 0.170 | 0.308 |  |  |  |  |
| **Contingency** | 0.540 | 0.590 | 0.314 |  |  |  |
| **Control** | 0.445 | 0.771 | 0.244 | 0.660 |  |  |
| **Causation** | 0.512 | 0.667 | 0.489 | 0.562 | 0.668 |  |

After the removal of the two items, we recalculated all previous indicators (see Tables 4 and 5). Individual reliability was met as the loads were within the thresholds defined above. The reliability of the construct was also met as the values of the composite reliability exceeded the limit of 0.8 and the ρA results were also higher than 0.7. Concerning convergent validity, we observed that all AVE values exceeded 0.5, so each item adequately represented a single construct.

**Table 4.** Convergent reliability and validity after eliminating items

| Constructs/Indicators | Loads | Composite Reliability | ρA | AVE |
|---|---|---|---|---|
| **Means** | | 0.868 | 0.773 | 0.686 |
| I use my personal knowledge and experience in the best possible way. | 0.829 | | | |
| I pursue those initiatives for which I have great motivation and interest. | 0.812 | | | |
| I pursue those initiatives for which I personally have the relevant competencies. | 0.843 | | | |
| **Partnership** | | 0.812 | 0.705 | 0.524 |
| When I work with others, I aim to ensure that gains and risks are shared fairly. | 0.610 | | | |
| I approach potential partners very early on in order to jointly co-create new things. | 0.808 | | | |
| I enter into relationships with partners who are willing to commit (e.g. invest time) from the onset. | 0.812 | | | |
| When new actors appear in my environment, I perceive them as potential partners. | 0.641 | | | |
| **Affordable Loss** | | 0.819 | 0.683 | 0.602 |
| I would only invest in a business what I can afford to lose. | 0.814 | | | |
| In a business, I would try to limit the potential loss of initiatives to an acceptable degree, although it could be that by investing more, I would finally obtain benefits. | 0.748 | | | |
| I would only invest in my business if the loss of the investment would not ruin the company, although it could be that by investing more, I would finally obtain benefits. | 0.764 | | | |
| **Contingency** | | 0.880 | 0.828 | 0.648 |
| I regard surprises to be new opportunities that I could take advantage of. | 0.780 | | | |
| I exploit contingencies as effectively as possible. | 0.888 | | | |
| When I have new information, I try to take advantage of it. | 0.743 | | | |
| I use setbacks as new opportunities to take advantage of. | 0.802 | | | |
| **Control** | | 0.835 | 0.744 | 0.561 |
| I attempt to shape the environment I operate in. | 0.810 | | | |
| I attempt to proactively design my environment with others. | 0.816 | | | |
| In a business, we must attempt to create with others new needs for the market. | 0.603 | | | |
| I attempt to influence trends. | 0.748 | | | |
| **Causation** | | 0.834 | 0.763 | 0.504 |
| I usually design a long-term plan to organise myself in my tasks. | 0.743 | | | |
| I prefer to have predetermined goals and to strive to achieve the results of these goals. | 0.785 | | | |
| I analyse long-run opportunities and select what I think would provide the best returns. | 0.757 | | | |
| I usually implement control processes to make sure I meet the objectives. | 0.637 | | | |
| I think my relationships with those who can influence my future should be long term and goal oriented. | 0.612 | | | |

Note: ρA: Dijkstra–Henseler; AVE: Average extracted variance

Finally, after the elimination of the two items, it was verified that discriminant validity still existed since the values of the HTMT ratio between the different variables were less than 0.85 (see Table 5).

Table 5. Discriminant validity after elimination of items

|  | Means | Partnership | Affordable Loss | Contingency | Control | Causation |
|---|---|---|---|---|---|---|
| **Means** |  |  |  |  |  |  |
| **Partnership** | 0.713 |  |  |  |  |  |
| **Affordable Loss** | 0.170 | 0.308 |  |  |  |  |
| **Contingency** | 0.540 | 0.590 | 0.314 |  |  |  |
| **Control** | 0.445 | 0.771 | 0.244 | 0.660 |  |  |
| **Causation** | 0.500 | 0.665 | 0.406 | 0.571 | 0.637 |  |

As indicated above, this research aims to develop and validate a measurement instrument that assesses the propensity of individuals towards effectual or causal behaviour when they have not yet created a company but could potentially do so. We designed a 25-item questionnaire, adapted from the work of Werhahn et al. (2015), Chandler et al. (2011), and Gabrielsson and Politis (2011), that measured six variables — five related to the effectual propensity of the individual (means, partnership, affordable losses, contingency, and control) and one related to causal propensity. After analysing the individual reliability, reliability of the construct, discriminant validity, and convergent validity of these constructs, we obtained a scale with 23 reliable and valid items (see Table 4). From the initial scale of 25 items, we eliminated only two items from the causal propensity construct to increase the explained variance and improve convergent validity. The items eliminated were: "I try to avoid uncertain situations to the greatest possible extent" and "When I set goals to achieve, I analyse my competitors in depth".

**DISCUSSION**

The results obtained show how an adequate adaptation of the scales proposed by Werhahn et al. (2015), Chandler et al. (2011), and Gabrielsson and Politis (2011) can be useful to measure the effectual and causal propensity of individuals who have not yet created a business.

This new scale has practical implications for management and the literature on entrepreneurial intention. Regarding the theoretical implications, this scale provides added value for entrepreneurship research interested in analysing the previous and initial stages of the entrepreneurial process. In this regard, the scale developed can contribute to the literature on entrepreneurial intention (Lechuga Sancho, Martín-Navarro, & Ramos-Rodríguez, 2020; Ramos-Rodríguez, Medina-Garrido, & Ruiz-Navarro, 2019) and opportunity recognition (Ramos-Rodríguez, Medina-Garrido, Lorenzo-Gómez, & Ruiz-Navarro, 2010), given the different behaviours that entrepreneurs might adopt at these stages of the entrepreneurial process depending on their effectual or causal orientation. This work offers a measurement instrument that fills a gap in the academic literature, as we have not found any scale that measures the propensity of individuals towards effectual or causal behaviour before starting a business. The results we have obtained fill this gap and represent an advance over previous analyses in the literature. This scale will allow other researchers to assess the effectual orientation of individuals who have not yet started a business in different contexts and populations, as its reliability and validity provide confidence for any future academic research design. In this way, this scale can contribute to the construction of more robust theories of entrepreneurial intention with effectual versus causal orientation.

In terms of practical implications, the measurement of effectual propensity will also be of practical use for those public and private agents with the function of promoting entrepreneurship since they will be able to pre-evaluate potential entrepreneurs and orient them towards a better use and fit to the environment of effectual and causal logics (Futterer et al., 2018). This scale will also be useful for the entrepreneurs and managers of already consolidated companies, who will be able to evaluate themselves and the members of their companies in order to implement the strategic orientation they consider most appropriate (Werhahn et al., 2015). Furthermore, effectual entrepreneurs who wish to assess the effectual propensity of their potential partners and workers may also use this scale of measurement before deciding to rely on them.

This tool will be particularly useful for the education system. Those academic institutions interested in developing the entrepreneurial potential of their students (Nowiński, Haddoud, Lančarič, Egerová, & Czeglédi, 2019; Padilla-Angulo, 2019; Ramos-Rodríguez et al., 2019) will have a validated measurement instrument with which to evaluate the initial propensity of their students towards an effectual or causal behaviour. This starting point will be useful when establishing the content of the training offered, allowing for the development of the logic (effectual versus causal) that is most innate to the students. Once academic institutions have detected and developed the dominant logic of each student, it is also advisable to train them in their non-dominant logic so that they can discern, apply, and combine both logics in order to improve their chances of success according to their entrepreneurial context (Futterer et al., 2018).

**CONCLUSIONS**

Effectual and causal logics suggest that entrepreneurs demonstrate different types of behaviour when they create and manage their companies and in their interactions with the environment (Sarasvathy, 2001). The literature on entrepreneurship has extensively studied these two types of behaviours (Brettel et al., 2012; Guo et al., 2016; Schmidt & Heidenreich, 2014). However, we have not found any studies that measure the individual's propensity towards these logics for the potential entrepreneur. Given the gap detected, the study of the propensity towards an effectual or causal behaviour supposes high added value to the literature on entrepreneurship.

In this sense, this research aims to develop and validate a measurement instrument that assesses an individual's propensity towards effectual or causal behaviour before they have started a company. To achieve this objective, we designed a questionnaire adapted from the works of Werhahn et al. (2015), Chandler et al. (2011), and Gabrielsson and Politis (2011). To check the reliability and validity of the items included in the questionnaire, a sample of final-year university students in the business administration programmes at the University of Cadiz and the University of Seville (Spain) was analysed, obtaining 230 valid responses. After analysing the individual reliability, construct reliability, discriminant validity, and convergent validity, we obtained a scale with 23 valid and reliable items (see Table 4).

This scale has important implications for the literature on potential entrepreneurship and entrepreneurial intention that is interested in the stages prior to the creation of a company. This work offers a measurement instrument that fills a gap in the literature regarding the measurement of individuals' propensity towards effectual or causal behaviour before starting a business. Likewise, the measurement of effectual propensity will be of practical use for: (1) entrepreneurship promoters who must guide the behaviour of entrepreneurs in a manner consistent with the environmental context (Futterer et al., 2018); (2) entrepreneurs and managers of consolidated companies who wish to implement a specific strategic orientation in their companies (Werhahn et al., 2015); (3) entrepreneurs who wish to measure the effectual propensity of their potential partners and workers; and (4) academic institutions interested in developing and orienting the entrepreneurial potential of their students.

However, some limitations must be kept in mind when interpreting these results. More empirical work is needed to increase the validity and generalisation of the established measures. Another limitation to consider is that we have used a sample with students from two universities in the same country. Future research should contrast this scale in different segments of the population in other countries and cultures. Another future line of research that would also fill a gap in the literature on entrepreneurship is the study of the moderating role of "effectual propensity" and "causal propensity" in entrepreneurial intentions (Arranz, Arroyabe, & Fdez. de Arroyabe, 2019; Dutta et al., 2015; Jeger, Sušanj, & Mijoč, 2014; Nowiński et al., 2019; Padilla-Angulo, 2019; Valliere, 2014, 2015).

**ACKNOWLEDGEMENTS**

This research has been partially granted by INDESS (Research Universitary Institute for Sustainable Social Development), Universidad de Cádiz, Spain.